\documentclass[11pt,dvips]{article}

\usepackage{epsfig,times,amssymb} 
\usepackage{picinpar}
\usepackage{wrapfig}
\usepackage{floatflt}
\makeatletter
\renewcommand{\section}{\@startsection{section}{1}{0in}
        {0.4\baselineskip}{0.1\baselineskip}{\Large\bf}}
\renewcommand{\subsection}{\@startsection{subsection}{2}{0in}
        {0.25\baselineskip}{-\baselineskip}{\large\bf}}
\renewcommand{\subsubsection}{\@startsection{subsubsection}{3}{0in}
        {0.1\baselineskip}{-\baselineskip}{\normalsize\bf}}
\makeatother
\begin{document}

%
%  Session and Paper Code:
%\thispagestyle{myheadings}
%
%  ***INSTRUCTIONS:***  Replace `OG 9.9.9' in the command argument below
%                       with your assigned session and paper code:
%\markright{HE 6.3.06}
%
\makeatletter\newcommand{\ps@icrc}{
\renewcommand{\@oddhead}{\slshape{HE.6.3.06}\hfil}}
\makeatother\thispagestyle{icrc}

\begin{center}
%
%  ***INSTRUCTIONS:***  Replace `Instructions for Preparation of Manuscript'
%                       with your paper's title:
{\LARGE \bf Capabilities of an Underwater Detector as a Neutrino Telescope
and for the Neutrino Oscillation Search}
\end{center}

%  Author List:
\begin{center}
%
%  ***INSTRUCTIONS:***  Replace authors and addresses below with your own:
%
{\bf T. Montaruli$^{1}$ for the NEMO Collaboration}\\
{\it $^{1}$Istituto Nazionale di Fisica Nucleare, Bari, I-70126, Italy\\}
\end{center}

%  Abstract:
\begin{center}
{\large \bf Abstract\\}
\end{center}
\vspace{-0.5ex}
We report on the results of a Monte Carlo simulation study of a km$^{3}$ scale 
deep underwater Cherenkov detector aimed at detecting neutrinos 
of astrophysical origin.
This analysis has been undertaken as part of the NEMO R\&D project 
to develop such an experiment close to the Southern Italian coasts.
We have studied the reconstruction capabilities of various arrays of 
phototubes in order to determine the detector geometries which optimize
performance and cost.
We have also investigated the possibility of designing a detector 
with characteristics suited to an experiment on atmospheric neutrino 
oscillations. 
%Capabilities and limits of this idea are discussed.
%
%  Leave this line skip in place:
\vspace{1ex}
%%%%%%%%%%%%%%%%%%%%%%%%%%%%%%%%%%%%%%%%%%%%%%%%%%%%%%%%%%%%%%%%%%%%%%%%%
\section{A km$^{3}$ Detector for Neutrino Astronomy:}
\begin{figwindow}[1,r,%
{\mbox{                  
\epsfig{file=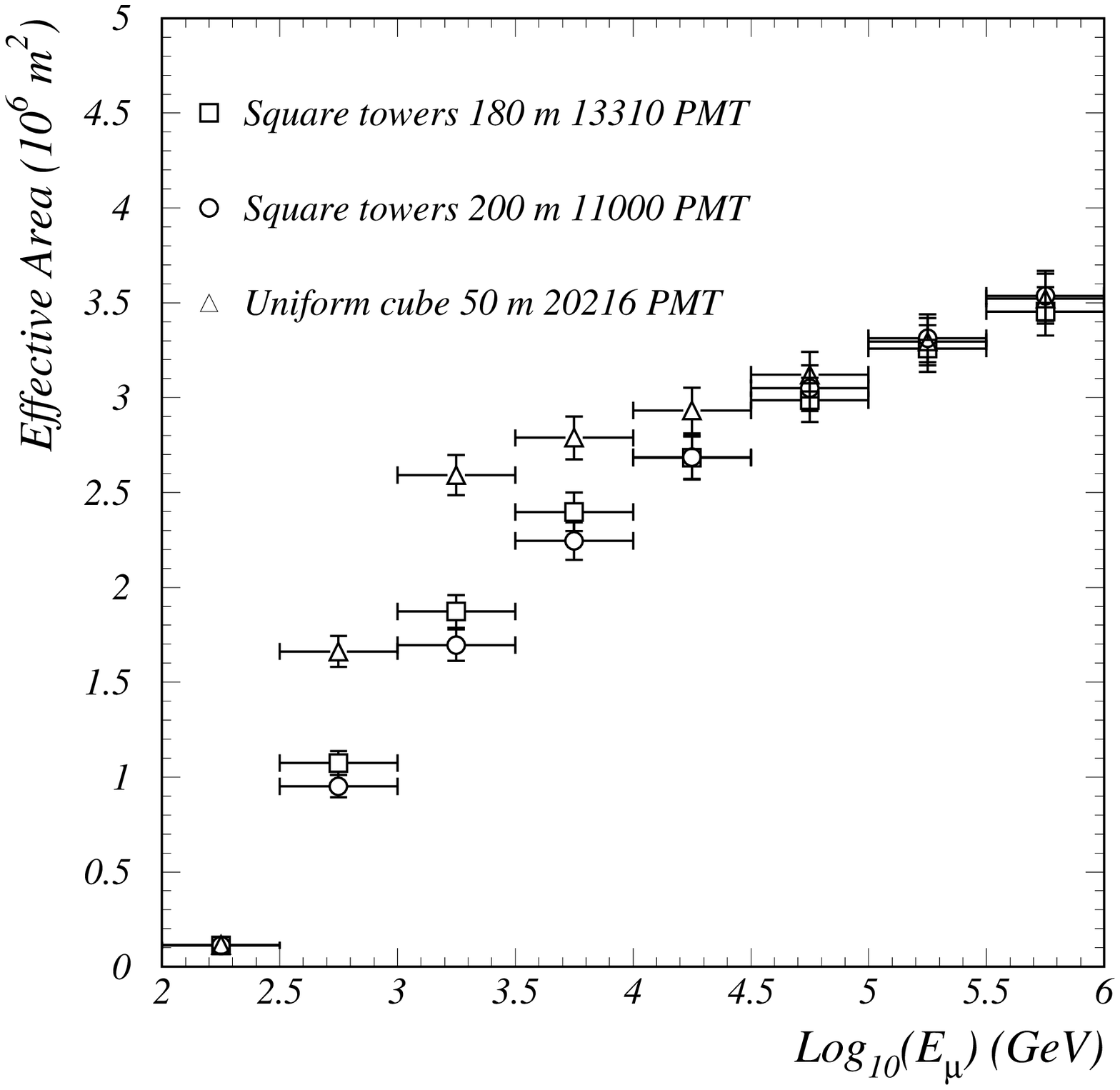,width=8.cm,height=8.5cm}
}},%
{Effective area vs muon energy 
for 3 PMT arrays: a parallelepiped with distance between PMTs of 50 m; 
2 square tower configurations with distance between centers of towers of 
180 m and 200 m and between PMTs in each tower of 30 m. Total numbers of 
PMTs are indicated for each array. The error bars are due to the Monte Carlo 
statistics.}]
The discovery of high energy neutrinos ($E_{\nu} > 100$ MeV) 
of astrophysical origin would open the $\nu$-astronomy field    
complementary to the well established $\gamma$-ray astronomy one.
Neutrinos are subject to less absorption than photons and therefore can
bring information 
on the deep interior of sources. The observation of tens of TeV $\gamma$-ray 
emitters (Thompson et al., 1995, Krennrich et al., 1998)
reinforces the possibility of existence of ``beam dump sources'',
accelerated proton beams interacting with gas of matter or photons.
Neutrino production is expected from $\pi^{\pm}$ decay in analogy with 
photons arising from $\pi^{0}$ decay.
Among the possible sources are active galactic nuclei (AGNs) 
made of a black hole 
and a surrounding accretion disk, binaries where a non compact 
companion transfers mass to the compact one (neutron star or black hole) with 
consequent development of an accretion disk,
supernova remnants in which particles interact in 
the acceleration region and $\gamma$-ray bursts where a fraction 
of the kinetic energy of a relativistic fireball is converted by photo-meson
production into neutrinos. 
Gaisser, Halzen \& Stanev, 1995, 
estimate rates of about 3 upwardgoing $\nu$-induced muons/yr 
and 0.1-25 muons/yr with $E_{\mu} > 1$ TeV in a 0.1 km$^2$ detector 
from galactic and extragalactic sources, respectively. These estimates
are based on existing limits from air shower experiments on $\gamma$
emissions in the 100 TeV energy range and on the assumption that $\nu$ fluxes 
are expected to be
comparable to $\gamma$-ray ones, except for $\gamma$ absorption.
Diffuse fluxes from AGNs may vary 
of orders of magnitude according to different models.
The signal from $\nu$ sources is expected to dominate the atmospheric 
$\nu$ background at energies $\gtrsim 10$ TeV, 
since the differential spectral 
index of high energy atmospheric neutrinos ($\gamma \sim 3.7$) 
is larger than that expected for $\nu$ cosmic
accelerators ($\gamma \sim 2 - 2.5$). 
Given the predicted low fluxes, the required sensitivity calls for
km$^{3}$ scale $\nu$ telescopes, with the ability to measure
the energy and direction of $\nu$-induced muons, in order to reject 
the atmospheric neutrino background. Underwater 
Cherenkov arrays of phototubes (PMTs)
can satisfy these requirements and can be developed at depths of the order
of 3000 m where the atmospheric muon background is reduced with respect to 
surface by a factor of $\sim 10^{-6}$.  
%The PMT times provide the versus of flight muons to discriminate 
%upward-going muons induced by neutrinos having crossed 
%the Earth with respect to the residual downward-going atmospheric muons.
\end{figwindow}
NEMO (NEutrino subMarine Observatory) is an R\&D project of the Italian INFN
for a $\nu$ telescope to be deployed in the Mediterranean Sea near 
the Southern Italian coasts, where transparency and other water parameters 
are optimal (Capone et al., 1999).
We have investigated the response of various arrays of
phototubes using a ``fast'' Monte Carlo simulation of Cherenkov light emission 
by high energy muons and of the detector response. 
The speed of the simulation is an important feature
at this stage of the feasibility study because it allows us to easily
change various inputs, such as the geometry of the array, the 
PMT characteristics and the kinematics of the events.
Moreover, it allows a fast propagation of muons above 10 TeV to be performed.
The main limit of this code is that it transports 
muons but not hadronic or electromagnetic showers.
We will perform more accurate studies with a full 
GEANT-based (Brun et al., 1987) 
simulation that we are developing (Bottai, 1998), once the
structure of the detector is well defined.
The total energy loss due to Cherenkov emission  
is orders of magnitude less important than the total ionization for
relativistic muons, but it has the relevant feature that all 
photons are emitted at $\Theta_{c} \sim 41.2^{\circ}$ 
around the $\mu$ trajectory. 
The simulation uses parameterizations to describe the
energy losses of muons by ionization and by 
stochastic processes (pair production, bremsstrahlung and nuclear 
interactions). The Cherenkov light emitted by electromagnetic showers
is produced according to the parameterization in Belyaev et al., 1979.
Light is attenuated as the result of absorption, 
which affects the amplitude of signals, and of scattering, 
which affects both the signal amplitude and the 
time of arrival with consequent difficulties in track reconstruction.
In the simulation, an attenuation length of light of 55 m has been assumed, 
as due to absorption only. 
Scattering of light is not simulated, but this is a 
reasonable approximation since measurements of the scattering 
length show that it is $\gtrsim 70$ m in the Mediterranean sea 
and the fraction of backscattered light is quite small. 
We have simulated optical modules (OMs) made of couples of 
typical byalcali PMTs 
looking one upward and the other downward, with a photocadode diameter of 
15 inch, time resolution
of 2.5 ns and quantum efficiency of 0.25. We have assumed 
a detection threshold of 0.25 photoelectrons (PE).
The events are assumed to give a trigger if they hit at least 5 PMTs.
The reconstruction of events requires in fact the determination of 5 parameters
(the zenith and azimuth angles and the coordinates of the  
``pseudo-vertex'', i.e. the position of the particle 
when the light hits the first PMT).
We perform it by means of a minimization procedure
of the $\chi^{2}$ between expected and measured arrival times of photons.
We are studying how the charge information can be used. 

In order to minimize the cost of $km^3$-size detectors,
a structure made of towers of strings of PMTs can be a better solution  
than uniformly spaced arrays. 
We have assumed 300 m high arrays.
% (which are more feasible in order 
%to be deployed, to avoid problems of gradients of currents and
%all possible differences between top and bottom).
Since we are interested in determining the response of the detector to
high energy $\nu$-induced muons produced by astrophysical sources,
we consider a relevant parameter the effective area (the area
including the reconstruction efficiency) as a function of  
energy for muons coming from outside the ``horizon'' of the detector, i.e.
a closed surface which contains the detector at a distance of 
2 attenuation lengths. 
In Fig. 1 we compare the effective areas 
of a cubic array of 50 m spaced PMTs (20,216 PMTs) with 2 
arrays made of square towers with their centers separated by 180 m and 200
m (13,310 and 11,000 PMTs, respectively). 
The square towers are made of 5 strings, 4 of which are located at the 
vertices and at a distance of 15 m from the one in the centre. 
The PMTs are vertically spaced by 30 m. It turns out from 
Fig. 1 that the areas
of the detectors are comparable; nevertheless the uniform array 
performs better in track reconstruction. 
%Anyway it should be considered that tracking is not yet 
%optimized on the structure.
%Using the model in Stecker et al, 1991, we find 7957 events/yr and 8744 
%events/year in the parallelepiped array and the tower configuration with
%distance 180 m.
%%%%%%%%%%%%%%%%%%%%%%%%%%%%%%%%%%%%%%%%%%%%%%%%%%%%%%%%%%%%%%%%%%%%%
\section{Neutrino Oscillations in an Underwater Detector:}
We have investigated the capabilities of an underwater array of determining the
oscillation parameters in the region suggested by current atmospheric 
$\nu$ experiments (Ambrosio et al., 1998; Fukuda et al., 1998), 
which in the $\nu_{\mu} \leftrightarrow \nu_{\tau}$ vacuum oscillation 
scenario is maximum mixing and $\Delta m^{2} \sim 3 \cdot 10^{-3}$ eV$^{2}$.
The oscillation probability is a function of the ratio of the 
baseline $L$ and of the neutrino energy $E_{\nu}$. 
Atmospheric neutrinos offer the 
possibility to explore a wide range in $L/E_{\nu}$ due to the
variation of the baseline with the zenith angle $\theta$
between $L \sim 20$ km for downward-going neutrinos and
$L \sim 2 R_{\oplus} \sim 12800$ km for
upward-going neutrinos.
Hence sensitivities to $\Delta m^{2} \sim 10^{-4}$ eV$^{2}$ can be achieved.
The measurement of the number of upward to downward events as a 
function of the ratio of the baseline to the 
energy, $N_{up}(L/E_{\nu})/N_{down}(L^{'}/E_{\nu})$, with 
$L^{'} = L(\pi - \theta)$, can provide the modulated pattern of 
the survival oscillation probability (Curioni et al., 1998). On the
other hand, if neutrinos do not oscillate the ratio is expected to be 1 
due to the up-down symmetry of $\nu$ fluxes above a few GeV. The 
possibility to measure downward-going events allows the normalization of
calculations to a ``reference'' flux not affected by oscillations.
As a matter of fact, if $E_{\nu} \sim 1$ GeV and $\Delta m^{2} \lesssim 
10^{-2}$ eV$^{2}$, downward neutrinos are not affected by oscillations, 
while upward neutrinos are reduced since $L/E_{\nu}$ ranges up to 10$^{4}$
km/GeV.

An underwater Cherenkov detector can provide 
indications of the neutrino direction and 
energy through the measurement of the induced muons.
The muon energy can be derived from the range of almost vertical muons
which do not leave the array (Moscoso et al., 1998).
We have defined the following variable to represent the vertical muon range:
the maximum difference of the heights of the highest and lowest hit PMTs 
($z_{max}-z_{min})_{max}$ among those calculated on each string. 
In the following we will assume that no information is obtained from the
hadronic part of the $\nu$ charged current (CC) interaction due to the 
insufficient granularity of the detector. In this 
approximation, a smearing will result in the oscillation pattern
due to the angle between the muon and the neutrino 
($\langle \Theta_{\nu\mu} \rangle = 6^{\circ}$ for the selected sample
of this analysis)
and to the energy transferred to the hadronic products.    
Nevertheless, we will show that the oscillation pattern can be significant
enough to discriminate the no-oscillation hypothesis and to
have some indication on the $\Delta m^{2}$ value.
After studying various configurations of dense arrays of PMTs we have
chosen the following configuration 
which reconstructs best tracks in the vertical 
region: 7 strings, with 1 in the center and 6 at the vertices of a hexagon
of 15 m side. OMs
are made of 2 PMTs looking upward and downward
and they are vertically spaced by 3 m; the strings are shifted in the
vertical direction with respect to one another by 1 m, in such a way that 
PMTs do not lie in the same horizontal planes. The total number of PMTs 
is 1414.
We find that increasing the vertical distance between OMs
causes a dramatic decrease in the statistics of events satisfying
the requirements of the analysis ($\sim 40\%$ for 6 m).
Since PMT arrays do not have a volume determined by 
active elements, we have generated 20.4 million events in a much larger 
volume than the tower dimensions 
(a cylinder of 72 m radius and 418 m high). 
We find that the effective volume of this tower of strings is about 
constant for $E_{\mu} \gtrsim 2$ GeV and its value is 
$2 \cdot 10^{6}$ m$^{3}$.  
Atmospheric $\nu$ interactions have been simulated by means of 
the NUIN generator
(Lipari et al., 1995) for $E_{\nu} \ge 1$ GeV with statistics 
corresponding to 34.4 yr of running time. We have used the Bartol group
$\nu$ flux, including recent calculations of geomagnetic cut-offs 
(Gaisser et al., 1988). 
The generator provides the kinematics of $\nu$-CC interactions for 
all flavors with particular attention to the processes which dominate at 
low energies (i.e. quasi-elastic scattering
and 1 $\pi$ production). We have used the parton structure functions 
GRV(94) (Gl\"uck, Reya \& Vogt, 1995) for deep inelastic scattering.
$\nu_{e}$ induced interactions have not yet been considered.
We have neglected the following sources of noise:
$\beta$-decay of $^{40}K$ and consequent light emission by 
electrons with a rate of the order of 50 kHz at the level of 1 PE signals
(this background should not represent a problem due to the requirement of
coincident measurement of 5 PMTs and to the applied filter that requires
that the coincidence happens in a time window compatible with light
propagation); bioluminescence bursts due to living organisms; high energy 
atmospheric muons whose vertical intensity at
3000 m depth is $\sim 10^{-8}$ cm$^{-2}$ s$^{-1}$ sr$^{-1}$. 
The sample of events is selected by 
requiring $|\cos\theta| > 0.94$, at least 1 
PMT hit on each string (this cut notably improves the quality of 
reconstruction), by imposing some 
containment cuts (-135 m $< z_{min} < z_{max} <$ 135 m,
the $x$ and $y$ coordinates of the reconstructed vertex
being inside a circle of radius 70 m and by requiring
-135 m $< |z| <$ 135 m) and $(z_{max}-z_{min})_{max} > 18$ m. The surviving
events are 750 yr$^{-1}$ for no-oscillations.
The angular separation $\Delta \Omega$ 
between the simulated and reconstructed $\mu$s is distributed 
with average angle of 3$^{\circ}$ and RMS of 
6.6$^{\circ}$. Only 1$\%$ of the selected events are badly reconstructed
with $\Delta \Omega > 20^{\circ}$. 
In Fig. 2(a) we show $(z_{max}-z_{min})_{max}$ vs the simulated $\mu$ energy.
The superimposed points forming a line show the ``true'' range of the $\mu$s.
We are working on an analytical correction to take into account
the $\mu$ track small angle with respect to the vertical in order to  
improve the energy resolution of the detector (i.e. the width of the
distribution around the ``true'' range). 
First estimates of the background due to atmospheric $\mu$s indicate that
even considering 6 planes of anticoincidence at the top of the
tower it may be necessary to reduce the fiducial volume of the detector. 
Better reconstruction algorithms may also help.
This requires the inclusion of the hadronic part of the interaction.
Even if preliminary, the results of this work are shown in Fig. 2(b) which 
represents the upward/downward muon ratio as a function of  
$(z_{max}-z_{min})_{max}$ for 5.7 yr of the tower running. As can be seen,
the dips of the curves obtained for $\Delta m^{2} = 1 \cdot
10^{-3}$, $2.5 \cdot 10^{-3}$ eV$^{2}$ are representative of the minima 
in the oscillation survival probability, while for $\Delta m^{2} = 5 \cdot
10^{-2}$  eV$^{2}$ the resolution of the experiment is such that
the fast oscillations are averaged at 1/2. 
Even if the tracking efficiency can be improved, more than 1 tower may be
needed in order to reach competitive statistics for this kind of experiment.
%\begin{figure}             
\begin{tabular}{cc}
\epsfig{file=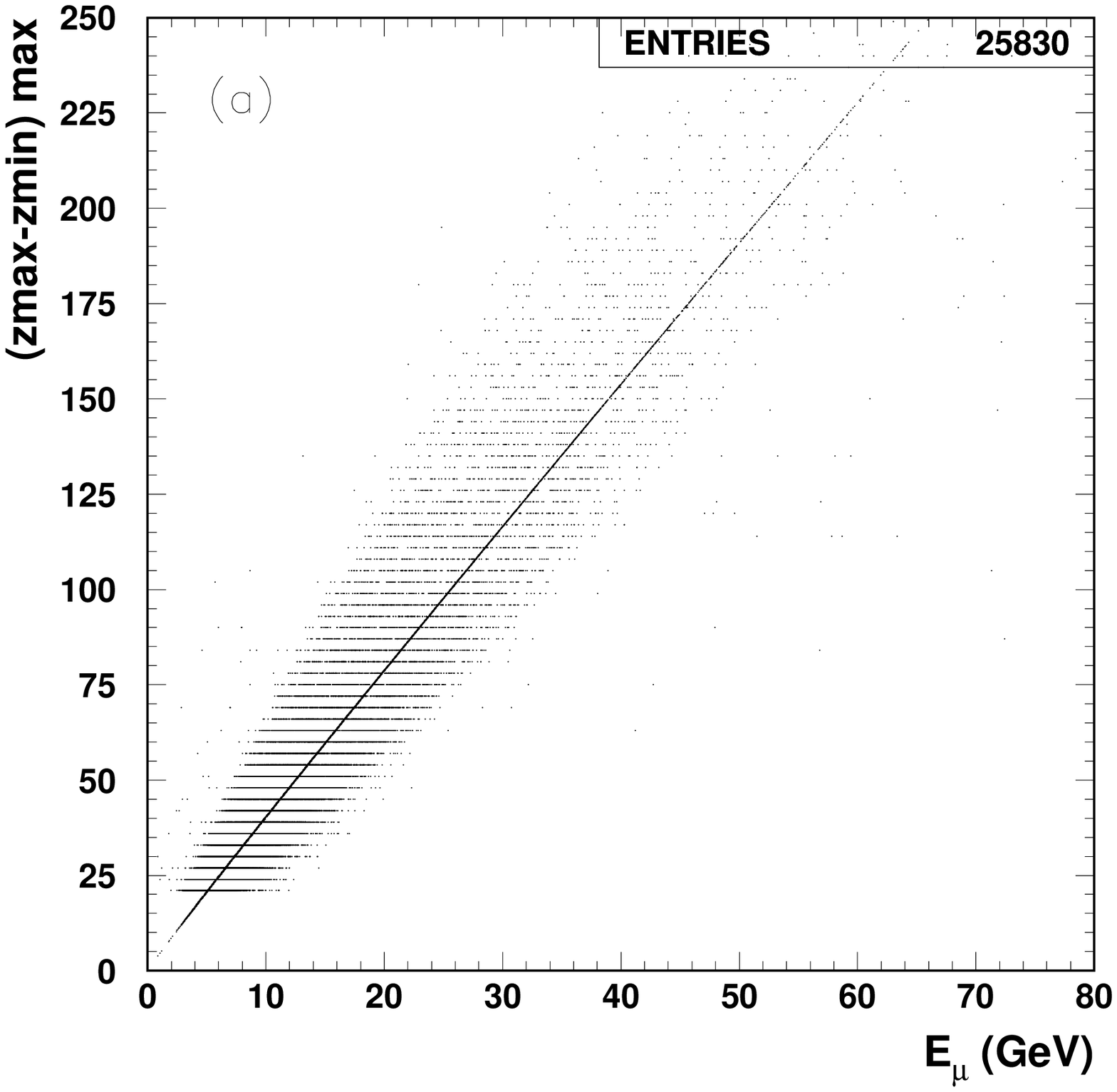,width=8.cm,height=8.5cm} &
\epsfig{file=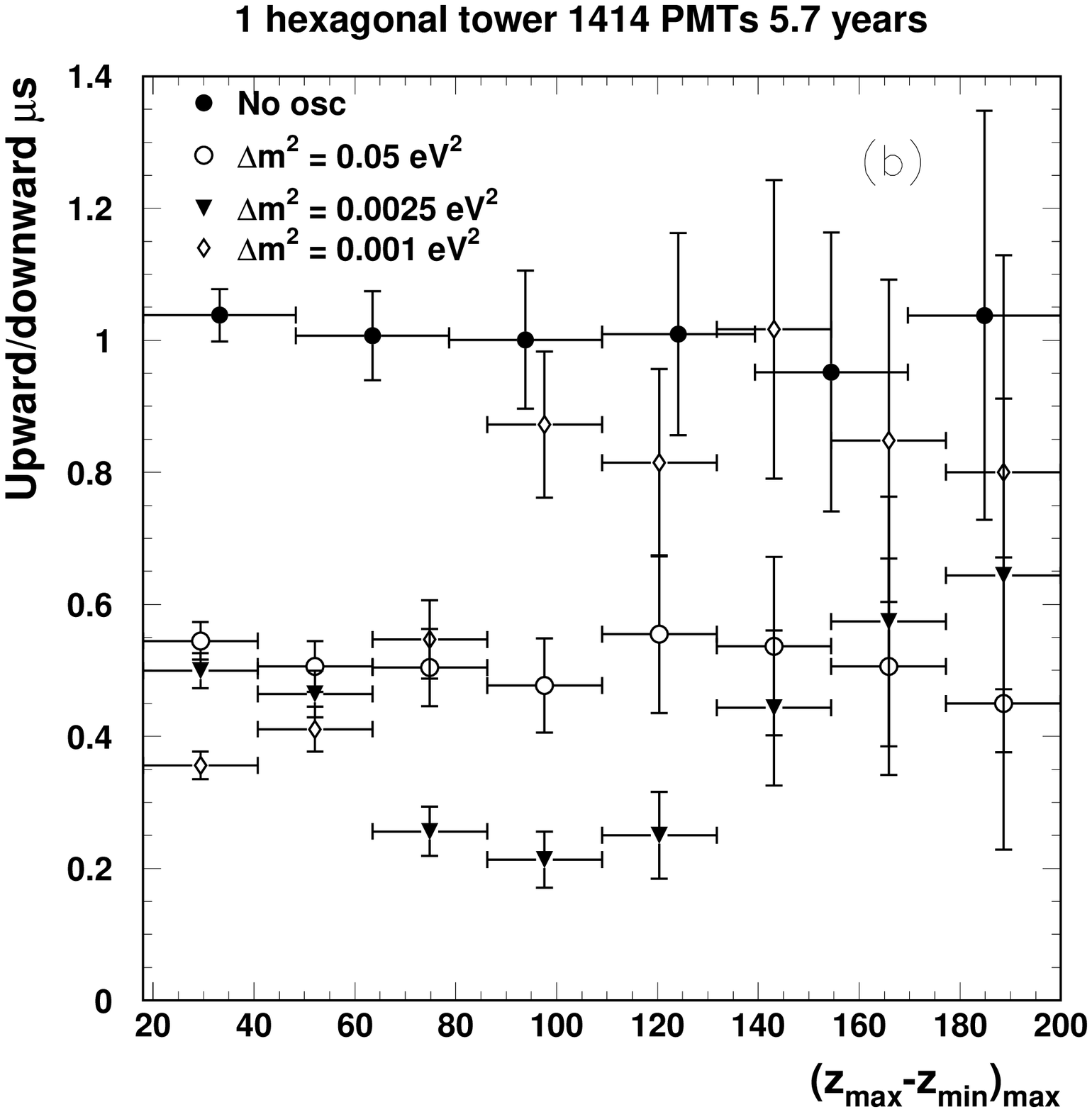,,width=8.cm,height=8.5cm}
\end{tabular}
%\caption{
\textbf{Figure 2: } (a) Correlation between $(z_{max}-z_{min})_{max}$ 
and the simulated energy of almost vertical muons. The points forming a line
represent the ``true'' range of simulated $\mu$s. (b) Upward/downward muon 
ratio vs $(z_{max}-z_{min})_{max}$ for the no-oscillation case and 
$\Delta m^{2} = 1 \cdot 10^{-3}$, 
$2.5 \cdot 10^{-3}$, $5 \cdot 10^{-2}$ eV$^{2}$. 
%}
%\end{figure}
\vskip .5 cm
\vspace{1ex}
\begin{center}
{\Large\bf References}
\end{center}
%
%  ***INSTRUCTIONS:***  Enter your references alphabetically following the 
% format
%                       of the example citations below.
Ambrosio, M., et al., MACRO Collaboration, 1998, Phys. Lett. B434, 451\\
Belyaev,  A.A., Ivanenko \& I.P., Makarov, V.V., 1979, Yad. Fiz. 30, 178\\  
Bottai, S., 1998, to appear in Proc. of "Simulation and Analysis Methods
for Large Neutrino Detectors" , DESY, Zeuthen and DFF 325/09/1998\\
Brun, R., et al, 1987, CERN report DD/EE/84-1\\
Capone, A., et al., NEMO Collaboration, 1999, these Proc. $26^{th}$ ICRC 
(Salt Lake City, 1999), HE.6.3.04\\
Curioni, A., et al., 1998, hep-ph/9805249\\ 
Fukuda, Y., et al., Superkamiokande Collaboration, 1998, Phys. Rev. Lett. 81,
1562\\
Gaisser, et al., 1988, Phys. Rev. D38, 85; Lipari, P., et al., 
1998, T., Phys. Rev. D58, 073003\\
Gaisser, T.K., Halzen, F. \& Stanev, T., 1995, Phys. Rep. 258, 173\\
Gl\"uck, M., Reya, E. \& Vogt, A., 1995, Z. Phys. C67, 433\\
Courtesy of  Lipari, P.; 
see Lipari, P., Lusignoli, M. \& Sartogo, F., 1995, Phys. Rev. Lett. 74, 4384\\
Moscoso, L. et al., ANTARES Collaboration, to appear in 
Proc. of NOW98, 7-9 Sep. 1998, Amsterdam\\
%Stecker, F.W., et al., 1991, Phys. Lett. {\bf 66} 2697\\
Thompson, D.J., et al., 1995, Ap. J. S, 101, 259; Krennrich, F., et al., 
1998, Astrop. Phys. 8, 213.
\end{document}